\documentclass[preprint,aps]{revtex4}
\usepackage{mathrsfs}
\usepackage{graphicx}
\usepackage{multirow}
\usepackage{makecell}
\usepackage{booktabs}
\usepackage{bm}
\usepackage{bbm}
\begin{document}

\title{Observation of spin-polarized surface states in a nodal line semimetal SnTaS$_{2}$}

\author{Chunyao Song$^{1,2\sharp}$, Lulu Liu$^{3\sharp}$, Shengtao Cui$^{4\sharp}$, Jingjing Gao$^{5\sharp}$, Pengbo Song$^{1,2\sharp}$, Lei Jin$^{6,7}$, Wenjuan Zhao$^{8}$, Zhe Sun$^{4}$, Xiaoming Zhang$^{6,7}$, Lin Zhao$^{1,2,9}$, Xuan Luo$^{5}$, Yuping Sun$^{5,10,11}$, Youguo Shi$^{1,2,9}$, Haijun Zhang$^{3,10*}$, Guodong Liu$^{1,2,9*}$ and X. J. Zhou$^{1,2,9,12*}$ }

\affiliation{
	\\$^{1}$Beijing National Laboratory for Condensed Matter Physics, Institute of Physics, Chinese Academy of Sciences, Beijing 100190, China
	\\$^{2}$University of Chinese Academy of Sciences, Beijing 100049, China
	\\$^{3}$National Laboratory of Solid State Microstructures, School of Physics, Nanjing University, Nanjing 210093, China
	\\$^{4}$National Synchrotron Radiation Laboratory, University of Science and Technology of China, Hefei 230029, China
	\\$^{5}$Key Laboratory of Materials Physics, Institute of Solid State Physics, HFIPS, Chinese Academy of Sciences, Hefei 230031, China
	\\$^{6}$State Key Laboratory of Reliability and Intelligence of Electrical Equipment, Hebei University of Technology, Tianjin 300130, China
	\\$^{7}$School of Materials Science and Engineering, Hebei University of Technology, Tianjin 300130, China
	\\$^{8}$Elettra Sincrotrone Trieste, Trieste 34149, Italy
	\\$^{9}$Songshan Lake Materials Laboratory, Dongguan 523808, China
	\\$^{10}$Collaborative Innovation Center of Advanced Microstructures, Nanjing University, Nanjing 210093, China
	\\$^{11}$High Magnetic Field Laboratory, HFIPS, Chinese Academy of Sciences, Hefei 230031, China
	\\$^{12}$Beijing Academy of Quantum Information Sciences, Beijing 100193, China
	\\$^{\sharp}$These authors contributed equally to the present work.
	\\$^{*}$Corresponding authors:XJZhou@iphy.ac.cn, gdliu$\_$ARPES@iphy.ac.cn and zhanghj@nju.edu.cn
}

\date{\today}

\pacs{}

\begin{abstract}

The superconductor SnTaS$_{2}$ is theoretically predicted to be an intriguing topological nodal line semimetal without consideration of spin-orbit coupling. By carrying out angle-resolved photoemission (ARPES) and spin-resolved ARPES measurements combined with band structure calculations, we have provided a complete picture of the electronic structure and spin polarization property for the prominent surface states of SnTaS$_{2}$. The low-energy electronic states are dominated by surface states; two of them are from the S-terminated surface, while four of them are from the Sn-terminated surface. These give rise to interesting Fermi surface topology of SnTaS$_{2}$: three pockets located at $\bar\Gamma$, $\bar{M}$ and $\bar{K}$ for the S-terminated surface and two pockets surrounding $\bar\Gamma$ and $\bar{K}$ for the Sn-terminated surface. We further reveal that two surface states that cross the Fermi level are spin-polarized. Since SnTaS$_{2}$ is also a superconductor, our observations indicate that it may provide a new platform to explore topological superconductivity and other exotic properties.

\end{abstract}

\maketitle

Topological superconductivity has become one of the frontier topics in condensed matter physics. A topological superconductor can lead to the realization of Majorana fermions, which are their own antiparticles and obey non-Abelian statistics, holding the potential to applications in fault-tolerant quantum computation\cite{Review_TSC_2011_RMP_Zhang,Review_TSC_2017_RPP_Ando}. Many approaches have been proposed to search for topological superconductors. The first one is looking for intrinsic topological superconductors with odd-parity pairing symmetry. However, such materials are rare in nature, and the proposed candidates like Sr$_{2}$RuO$_{4}$\cite{Review_Sr2RuO4_2003_RMP_Maeno,Sr2RuO4_2013_PRL_Sato,Sr2RuO4_Exp_2019_Nature_Brown} and Cu$_{x}$Bi$_{2}$Se$_{3}$\cite{CuxBi2Se3_Theory_2010_PRL_Berg,CuxBi2Se3_Transport_2010_PRL_Cava,CuxBi2Se3_Pointcontact_2011_PRL_Ando,CuxBi2Se3_STM_2013_PRL_Stroscio,CuxBi2Se3_NMR_2016_NatPhy_Zheng} remain controversial. The second approach is to construct artificial hybrid systems like heterostructures of superconductors and topological insulators\cite{TSCProximityEffect_2008_PRL_Kane}, or quantum wires with strong spin-orbit coupling (SOC) embedded into a superconducting quantum interference device\cite{Heterostructure_1DQuanWireSC_2010_PRL_Sarma}, or a combination of magnetism and superconductivity\cite{Magnetic_2011_PRB_Beenakker,Magnetic_Exp_2014_Science_Ali}. This route, however, suffers from the complications of managing the interface and the limitation of the superconducting coherence length. To overcome these shortcomings, the latest advancement is to find topological superconductivity in a single material that is superconducting and coincidentally possesses topological surface states\cite{Review_TSC_2018_NSR_Hu}. These have resulted in a few candidates like FeSe$_{x}$Te$_{1-x}$\cite{FeSeTe_Theory_2016_PRL_Zhang,FeSeTe_ARPES_2018_Science_Shin,FeSeTe_STM_2018_Science_Gao} and Li(Fe,Co)As\cite{LiFeCoAs_ARPES_2019_NatPhy_Shin} that require fine-tuning in the chemical composition, and  LiFeOHFeSe\cite{LiFeOHFeSe_TSC_2018_PRX_Feng}, PbTaSe$_{2}$\cite{PbTaSe2_ARPES_2016_Natcom_Hasan}, 2M-WS$_{2}$\cite{WS2_Crystal_2019_AdvMater_Huang,WS2_STM_2019_NatPhys_Xue,WS2_ARPES_2021_NatCom_Chen} and TaSe$_{3}$\cite{TaSe3_ARPES_2020_Matter_Chen} that are stoichiometric.

Quite similar to the well-characterized topological superconductor PbTaSe$_{2}$ in the lattice structure, SnTaS$_{2}$ (T$_{c}$$\sim$3\,K) is found to be a topological semimetal and also a superconductor. Without SOC, a topological nodal line state is formed at the Brillouin zone corner surrounding the K point\cite{SnTaS2_Crystalstructure_2019_PRB_Teng,SnTaS2_ARPES_2020_PRB_Shen}. ARPES measurements provide evidence for such prediction from band calculation\cite{SnTaS2_ARPES_2020_PRB_Shen}. However, due to band crossing appearing above E$_{F}$, it was impossible to identify the Ta-derived bands around K and observe the important band inversion. This leads to difficulty in confirming the nodal line state. In this paper, we observed the clear and clean band crossing and relevant surface state with spin polarization. By performing high-resolution ARPES, spin-resolved ARPES, and first principle calculations, we identified the Dirac nodal lines in SnTaS$_{2}$. Two branches of the surface states are observed that cross the Fermi level and are split by the spin-orbit coupling. The spin-resolved ARPES measurements indicated that these surface states are spin-polarized. These results make SnTaS$_{2}$ a possible new candidate in realizing topological superconductivity.

High-quality single crystals of SnTaS$_{2}$ were grown by the chemical vapor transport method\cite{SnTaS2_Crystalstructure_2019_PRB_Teng} and were characterized by X-ray diffraction and magnetic measurement (Fig. S1 in Supplemental Material). ARPES measurements and spin-resolved ARPES measurements were performed at BL-13U in National Synchrotron Radiation Laboratory (NSRL, Hefei, China) equipped with a Scienta Omicron DA30L analyzer. The VLEED spin detector directly attached to the DA30L analyzer can detect two spin components. One component is in-plane, while the other is out-of-plane. The total energy and angular resolutions for ARPES measurements were set as $\sim$17 meV and $\sim$0.3 degrees, respectively. The total energy for the spin-resolved ARPES measurements was set as $\sim$30 meV. Two polarization geometries were used. In the linear horizontal (LH) polarization geometry, the electric field vector of the incident light is parallel to the horizontal plane. In contrast, in the linear vertical (LV) polarization geometry, the electric field vector is perpendicular to the horizontal plane. The SnTaS$_{2}$ single crystal samples were cleaved {\it in situ} and the (001) surface was measured at a temperature of 18\,K in ultrahigh vacuum better than 6.0$\times$10$^{-11}$ mbar. The structural optimizations and electronic property calculations were performed in the framework of density functional theory under the Perdew-Burke-Ernzerhof (PBE)\cite{Calculation_PBE_1996_PRL_Ernzerhof} of generalized gradient approximation (GGA)\cite{Calculation_GGA_1992_PRB_Carlos} as implemented in the VASP package\cite{Calculation_VASP_1996_CMS_Furthmiiller}. Pseudopotentials were employed within the scalar relativistic projector augmented wave (PAW)\cite{Calculation_PAW_1994_PRB_Blochl} method with 5s$^{2}$5p$^{2}$ valence electrons for Sn, 5d$^{3}$6p$^{2}$ valence electrons for Ta, and 3s$^{2}$3p$^{4}$ valence electrons for S. The plane-wave cutoff energy was set to be 420 eV. The optimized lattice parameters a = b = 3.319 $\AA$, c = 17.450 $\AA$ are employed in the calculations.

SnTaS$_{2}$ crystallizes in a nonsymmorphic hexagonal structure with a space group of P6$_{3}$/mmc (No.194). The Sn and TaS$_{2}$ layers stack alternatively along the c axis, as illustrated in Fig. 1a and 1b. We carried out extensive ARPES measurements on SnTaS$_{2}$. We found two kinds of cleaved sample surfaces that show different electronic structures. The measured Fermi surface and band structures on the first kind of surface are shown in Fig. S2 in Supplemental Material, while those measured on the second kind of surface are presented in Fig. 1 and 2. The observed electronic structures on the first kind of surface (Fig. S2) are similar to those reported before\cite{SnTaS2_ARPES_2020_PRB_Shen}. The measured electronic structures on the second kind of surface (Fig. 1 and 2) encompass all the features observed on the first kind of surface with additional features that were not reported before. As we will show below, the more complete electronic structures from the second kind of surface are more consistent with the band structure calculations. Therefore, we will mainly concentrate on the electronic structures on the second kind of surface.

Figure 1d and 1e show the measured Fermi surface mapping and constant energy contour at a binding energy of 0.3\,eV. Combined with the analysis of the measured band structures (Fig. 2), the measured Fermi surface of SnTaS$_{2}$ is determined and shown in Fig. 1f. Five Fermi surface sheets are observed. Two hole-like hexagon-shaped Fermi surface sheets ($\alpha$ and $\beta$ in Fig. 1f) are around the $\bar\Gamma$ point. An electron-like dog-bone-shaped Fermi surface sheet ($\gamma$ in Fig. 1f) is observed around the $\bar{M}$ point. Two triangle-shaped Fermi surface sheets, one electron-like ($\varepsilon$ in Fig. 1f) and the other hole-like ($\delta$ in Fig. 1f) are present around the $\bar{K}$ point. We note that the $\beta$ and $\varepsilon$ Fermi surface are additional sheets that are not observed in the measurements on the first kind of surface (Fig. S2)\cite{SnTaS2_ARPES_2020_PRB_Shen}.

Figure 2 shows the band structures of SnTaS$_{2}$ measured along high symmetry directions. Considering the photoemission matrix element effects, we also measured the band structures under different polarization geometries to get the complete electronic structure and analyze the orbital contributions, as shown in Fig. S3 in Supplemental Material. To understand the origin of the observed bands, we also plot the calculated band structures of SnTaS$_{2}$ both for the bulk and for the surface states. Fig. 2a shows the band structure measured along the $\bar\Gamma$-$\bar{K}$ direction. Fig. 2b shows the calculated band structure of SnTaS$_{2}$ along $\Gamma$-K. The surface states deduced from the slab calculations are shown as red lines in Fig. 2c. Fig. 2d shows the overall calculated band structures along $\Gamma$-K by combining both the bulk states and the surface states. By shifting the Fermi level of the calculated band structures upwards by 0.225\,eV, we find that the measured band structure in Fig. 2a shows an overall good agreement with the calculated band structures in Fig. 2d. A careful comparison between the measured bands in Fig. 2a and the calculated bands in Fig. 2d indicates that the measured band structures are mainly dominated by the surface states. As marked in Fig. 2a and 2d, four electronic states labeled SS1, SS2, SS3, and SS4, can be clearly identified. The SS1 and SS2 states show a clear band splitting. Since the band splitting occurs only for the surface states, as seen in Fig. 2c and Fig. 2d, these two states can be clearly attributed to the surface states. Since the expected bulk bands are rather weak, which are either invisible or mingle with the surface states, the observed SS3 and SS4 states are more likely dominated by the surface states, although how much bulk states may be mixed needs further investigation.

Figure 2e and 2f show the band structure measured along the $\bar{K}$-$\bar{M}$-$\bar{K}$ direction. Fig. 2i shows the overall calculated band structures along K-M-K by combining both the bulk states from Fig. 2g and the surface states from Fig. 2h. The measured band structures in Fig. 2e and 2f show an excellent agreement with the calculated band structures in Fig. 2i. Again, the measured band structures are dominated by the surface states. As marked in Fig. 2e and 2i, two surface states, labeled SS5 and SS6, can be clearly observed. The bulk bands are also rather weak and are hardly visible in the measured data. The band structures measured along $\bar\Gamma$-$\bar{K}$ and $\bar{K}$-$\bar{M}$-$\bar{K}$ using different photon energies (Fig. S4 in Supplemental Material) show little difference. This indicates a weak k$_{z}$ dependence of the observed bands, which further supports the nature of the dominant surface states. The good agreement between the measured and calculated bands, and the overlapping of some surface state bands with the bulk bands, make it possible to identify the Dirac nodes expected from the band structure calculations, as shown in Fig. S5 in Supplemental Material.

To further understand the origin of the observed surface states, we carried out slab calculations and decomposed the contribution from each layer in the slab. The calculated results are presented in Fig. 3b to 3g for a six-unit-cell-thick slab shown in Fig. 3a. In the slab, the first layer corresponds to the Sn-terminated surface, while the sixth layer corresponds to the S-terminated surface. As seen in Fig. 3b, for the first layer with Sn-termination, four main branches of surface states can be identified, labeled SS1, SS3, SS4, and SS6. For the sixth layer with S-termination, two main branches of surface states appear near the Fermi level, labeled SS2 and SS5. These calculated results indicate that the observed six branches of surface states in Fig. 2 originate from two kinds of surfaces with different terminations. These also explain the difference between the two kinds of sample surfaces we have measured. In the first kind of surface, the S-terminated regions dominated, and mainly the SS2 and SS5 surface states are observed (Fig. S2 in Supplemental Material)\cite{SnTaS2_ARPES_2020_PRB_Shen}. In the second kind of surface, the Sn-terminated and S-terminated regions are comparable, and all six surface states can be observed. These also reveal the origin of the observed five Fermi surface sheets in Fig. 1d and 1f. The $\alpha$, $\gamma$, and $\delta$ Fermi surface come from the SS2 and SS5 surface states from the S-terminated region, while the $\beta$, and $\varepsilon$ Fermi surface come from the SS3 and SS4 states from the Sn-terminated region. It deserves to point out that the SS4 state observed here is, in terms of shape, energy, and momentum position, quite similar to the predicted drumhead surface state from band calculation with no SOC included.
 
Now we come to determine the spin polarization of the surface states in SnTaS$_{2}$. Fig. 4 shows the measured results of the SS2 surface states observed along $\bar\Gamma$-$\bar{K}$ (Fig. 4a-4e) and the SS5 surface states observed along $\bar{K}$-$\bar{M}$-$\bar{K}$ (Fig. 4f-4j). Because of the SOC, both SS2 and SS5 split into two branches. The spin polarization is defined by three components along the X, Y, and Z directions, as shown in the inset of Fig. 4k. Each component is determined from a pair of energy distribution curves (EDCs) measured at a given momentum point under a specific configuration. Their difference corresponds to the spin polarization, and the sign of the EDC difference determines the spin orientation. For the SS2 surface state shown in Fig. 4a, both measurements at P1 (Fig. 4b) and P2 (Fig. 4c) indicate that the spin polarization along the Z direction is opposite on the two split bands, SS2U and SS2D. The spin polarization along the Y direction is also opposite on these two bands, as shown in Fig. 4d and 4e. These results clearly demonstrate that the two split bands of the SS2 surface states are spin-polarized, and they have opposite spin orientations.

For the Rashba-like SS5 surface state shown in Fig. 4f, at the P3 point, the two branches SS5L and SS5R degenerate at the $\bar{M}$ point. Therefore, the states near -0.5\,eV show no spin polarization both along the Z direction (Fig. 4g) and along the Y direction (Fig. 4i). At the P4 point where the SS5L and SS5R bands clearly separate, they are polarized both along Z (Fig. 4h) and Y (Fig. 4j) directions. Furthermore, the spin orientation on the two bands is opposite both along Z and Y directions. We note that the magnitude of the spin polarization for the SS5 surface state (Fig. 4h and 4j) appears to be much weaker than that for the SS2 surface state (Fig. 4c and 4e). To our best knowledge, it is the first time that the Sn-terminated surface states were clearly and comprehensively observed. We speculate that such states should originate from a strong surface electric field due to a perfect Sn-terminated cleaving surface. The surface electric field also contributes to the Rashba-like band splitting and spin polarization.

In summary, by carrying out ARPES and spin-resolved ARPES measurements combined with band structure calculations, we have provided a complete picture of the electronic structure and the spin-polarized surface state of SnTaS$_{2}$. The low-energy electronic states are dominated by surface states; two of them are from the S-terminated surface, while four of them are from the Sn-terminated surface. These give rise to interesting Fermi surface topology of SnTaS$_{2}$: three pockets located at $\bar\Gamma$, $\bar{M}$ and $\bar{K}$ for the S-terminated surface (Fig. 4l) and two pockets surrounding $\bar\Gamma$ and $\bar{K}$ for the Sn-terminated surface (Fig. 4m). We further reveal that the surface states SS2 and SS5, which cross the Fermi level, split due to the SOC and are spin polarized. Signatures of the Dirac nodes are identified. Since SnTaS$_{2}$ is superconducting, these intriguing observations make it a rich platform to explore topological superconductivity and other exotic properties.

\hspace*{\fill}

\vspace{3mm}

\noindent {\bf Acknowledgement}\\
This work is supported by the National Natural Science Foundation of China (Nos. 11974404, 12074181, 11922414, 11888101, 11674326 and 11774351), the National Key Research and Development Program of China (Nos. 2018YFA0305602, 2021YFA1401800 and 2021YFA1600201), the Strategic Priority Research Program (B) of the Chinese Academy of Sciences (Nos. XDB25000000), the Youth Innovation Promotion Association of CAS (No. 2017013), Synergetic Extreme Condition User Facility (SECUF), the Fundamental Research Funds for the Central Universities (Grant No. 020414380185), Natural Science Foundation of Jiangsu Province (No. BK20200007) and the Joint Funds of the National Natural Science Foundation of China and the Chinese Academy of Sciences’ Large-Scale Scientific Facility (Nos. U1832141, U1932217 and U2032215).

\vspace{3mm}

\noindent {\bf Author Contributions}\\

G.D.L., C.Y.S. and X.J.Z. proposed and designed the research. J.J.G., P.B.S., X.L., Y.P.S and Y.G.S. contributed to the SnTaS$_{2}$ crystal growth. L.L.L., L.J., X.M.Z. and H.J.Z. contributed to the band structure calculations and theoretical discussions. C.Y.S., S.T.C., G.D.L. and Z.S. contributed to the ARPES experiment at NSRL. C.Y.S., G.D.L. and X.J.Z. analyzed the data. C.Y.S., G.D.L. and X.J.Z. wrote the paper.  All authors participated in discussion and comment on the paper.

\newpage

\begin{figure*}[tbp]
	\begin{center}
		\includegraphics[width=1\columnwidth,angle=0]{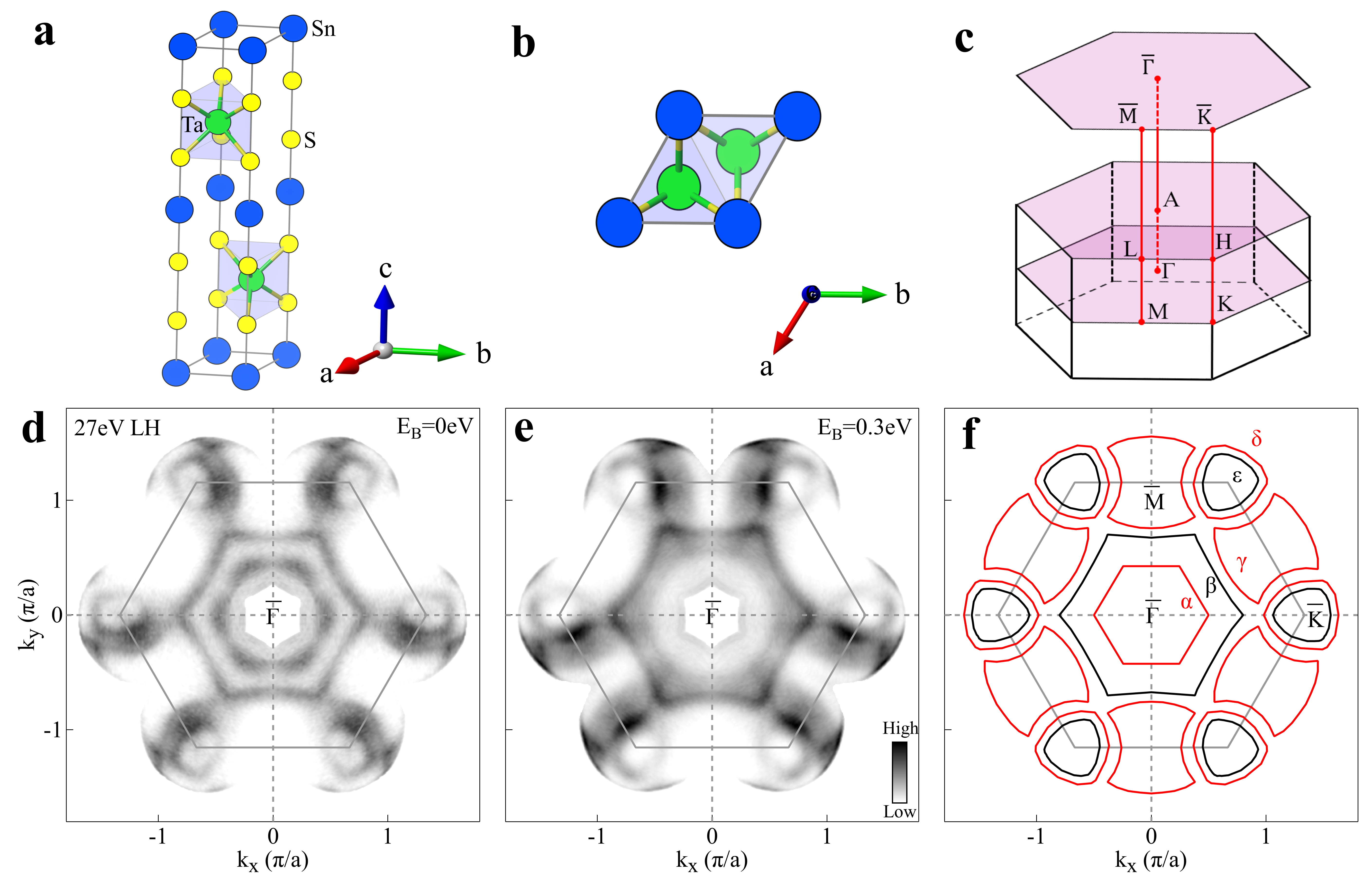}
	\end{center}
	\caption{\textbf{The measured Fermi surface of SnTaS$_{2}$.} (a) Crystal structure of SnTaS$_{2}$. It shows a layered structure formed by the alternative stacking of Sn and TaS$_{2}$ layers. (b) Top view of crystal structure. (c) Three-dimensional (3D) Brillouin Zone of SnTaS$_{2}$ and the projected two-dimensional (2D) Brillouin Zone. High symmetry points are marked. (d-e) Fermi surface mapping (d) and constant energy contour at the binding energy of 0.3\,eV (e) measured with a photon energy of 27\,eV. (f) Schematic Fermi surface of SnTaS$_{2}$ based on the measured Fermi surface (d) and related band structure analysis. The Fermi surface sheets originated from bulk, and surface states are drawn by black lines and red lines, respectively.
	} 
\end{figure*}

\begin{figure*}[tbp]
	\begin{center}
		\includegraphics[width=1\columnwidth,angle=0]{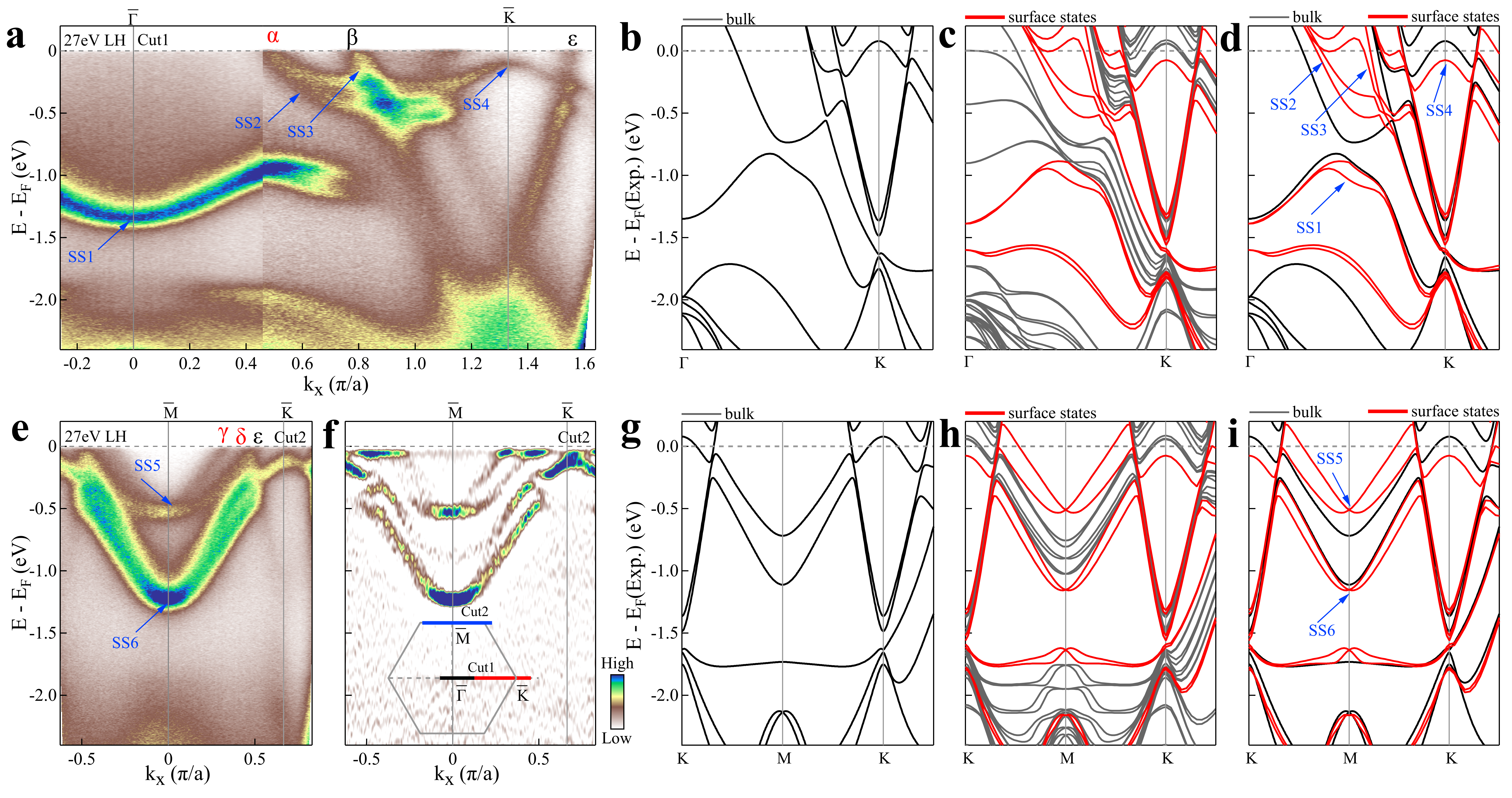}
	\end{center}
	\caption{\textbf{Band structures of SnTaS$_{2}$ measured along high symmetry directions and their comparison with band structure calculations.} (a) Band structure measured along $\bar{\Gamma}$-$\bar{K}$ with a photon energy of 27\,eV under the LH polarization geometry. The location of the momentum cut is marked by Cut1 in the inset of (f). (b) Calculated bulk band structures of SnTaS$_{2}$ with SOC along $\Gamma$-K. (c) The calculated band structures of SnTaS$_{2}$ with SOC along $\Gamma$-K for a six-unit-cell-thick slab. The surface states are marked by red curves. (d) Calculated band structures of SnTaS$_{2}$ along $\Gamma$-K combining the bulk bands in (b) and the surface states in (c). (e) Band structure measured along $\bar{K}-\bar{M}-\bar{K}$ with a photon energy of 27\,eV under the LH polarization geometry. The location of the momentum cut is marked by Cut2 in the inset of (f). (f) Corresponding second derivative image of (e) with respect to the energy. (g-i) Same as (b-d) but along K-M-K.
	}
\end{figure*}

\begin{figure*}[tbp]
	\begin{center}
		\includegraphics[width=1\columnwidth,angle=0]{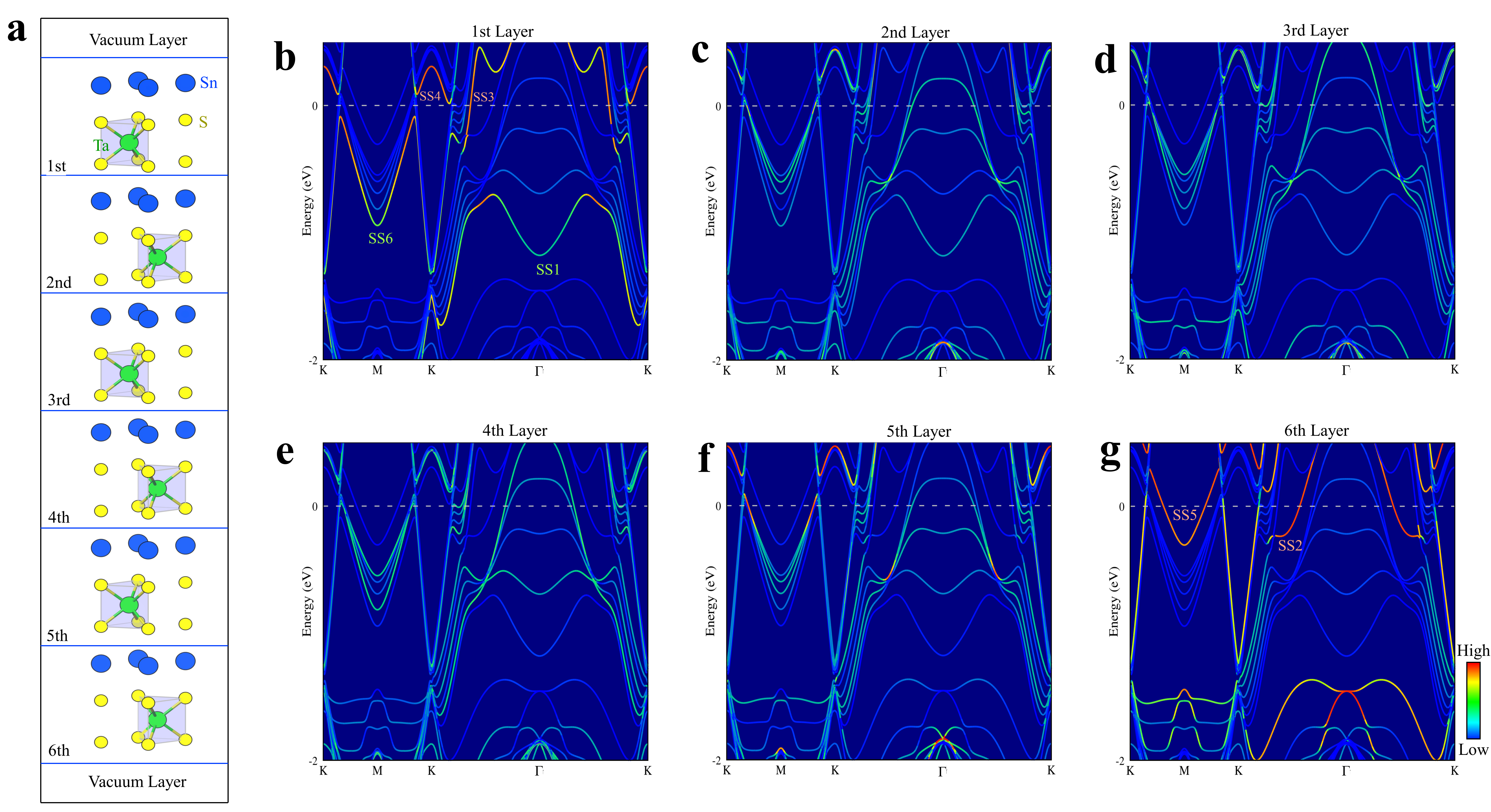}
	\end{center}
	\caption{\textbf{Decomposed contribution by different unit layer in the calculated band structures of SnTaS$_{2}$ for a six-unit-cell-thick slab.} For simplicity, the SOC is not considered in the calculations. (a) Structure of the six-unit-cell-thick slab used in the calculation. The first layer corresponds to the Sn-terminated surface, while the sixth layer corresponds to the S-terminated surface. (b-g) Contribution of band structures by different unit layers from the first layer (b) to the sixth layer (g). The surface states SS1, SS3, SS4, and SS6 come from the first layer and are marked in (b). The surface states SS2 and SS5 come from the sixth layer and are marked in (g).
	}
\end{figure*}

\begin{figure*}[tbp]
	\begin{center}
		\includegraphics[width=1\columnwidth,angle=0]{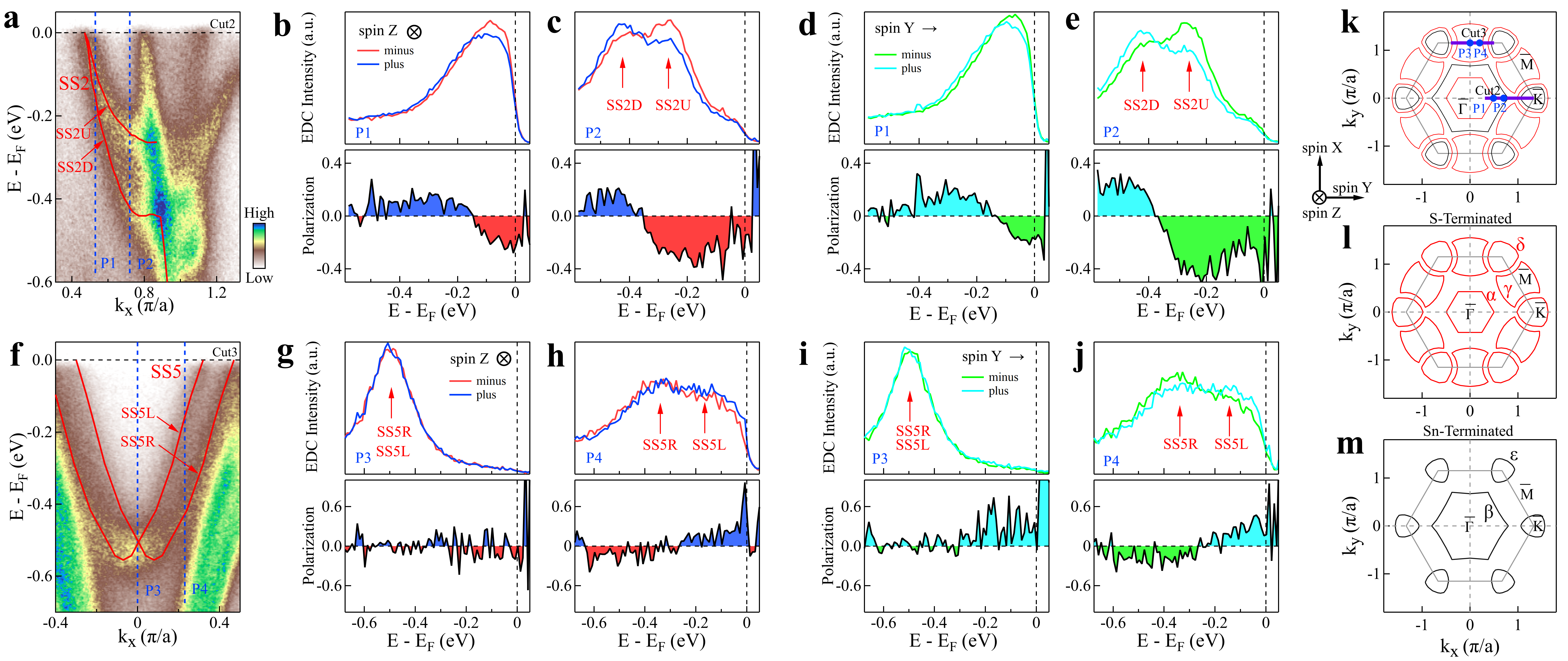}
	\end{center}
	\caption{\textbf{Spin polarization determination of the surface states in SnTaS$_{2}$} (a) Band structure measured along $\bar{\Gamma}$-$\bar{K}$. The location of the corresponding momentum Cut2 is marked by a purple line in (k). The SS2 surface state is observed that split into two branches labeled SS2U and SS2D. (b-c) Spin-resolved EDCs (upper panels) and the corresponding spin polarization curves (lower panels) for determining the Z component of the spin polarization of the momentum points P1 (b) and P2 (c). The locations of P1 and P2 are marked in (a) and also in (k). (d-e) Same as (b-c) but for determining the Y component of the spin polarization of the momentum points P1 (d) and P2 (e). (f) Same as (a), but along $\bar{K}-\bar{M}-\bar{K}$ as marked by Cut3 in (k). The SS5 surface state shows a Rashba-like splitting that consists of two branches labeled SS5L and SS5R. (g-j) Same as (b-e) but for the momentum points P3 and P4 marked in (f) and also in (k). (k) Schematic Fermi surface of SnTaS$_{2}$. Relative to the sample orientation set by the first Brillouin zone, the three spin components (X, Y, and Z) are defined in the bottom-left inset. (l) Schematic Fermi surface of the S-terminated surface of SnTaS$_{2}$ consisting of $\alpha$, $\gamma$ and $\delta$ sheets. (m) Fermi surface of the Sn-terminated surface with two sheets of $\beta$ and $\varepsilon$.
	}
\end{figure*}

\end{document}